\documentclass[aps,prd,preprint,preprintnumbers,showpacs,nofootinbib]{revtex4}

\usepackage{graphicx}

\usepackage[utf8]{inputenc} 

\usepackage{amsmath}
\usepackage{amssymb}
\usepackage{float}
\usepackage{comment}

\usepackage{soul}

\usepackage[usenames,dvipsnames]{color}

\definecolor{mightnightblue}{RGB}{25,25,112}

\definecolor{brown}{rgb}{0.59, 0.29, 0.0}

\def\21{$\mathrm{SU(2)_L \otimes U(1)_Y}$}

\newcommand{\AddrAHEP}{AHEP Group, Institut de F\'{i}sica Corpuscular --
  C.S.I.C./Universitat de Val\`{e}ncia, Parc Cientific de Paterna.\\
  C/Catedratico Jos\'e Beltr\'an, 2 E-46980 Paterna (Val\`{e}ncia) - SPAIN}

\newcommand{\Cinvestav}{Departamento de F\'{\i}sica, Centro de
  Investigaci{\'o}n y de Estudios Avanzados del IPN\\ Apdo. Postal
  14-740 07000 Mexico, DF, Mexico}
  
  \newcommand{\campinas}{Instituto de F\'isica Gleb Wataghin - UNICAMP, {13083-859}, Campinas SP, Brazil}

\usepackage{hyperref}
 \hypersetup{
     colorlinks=true,
     linkcolor= Black,
     citecolor=mightnightblue,
     urlcolor=mightnightblue
     } 

\begin{document}


\title{Exploring the Potential of Short-Baseline Physics at Fermilab}

\author{O. G. Miranda~$^1$}\email{omr@fis.cinvestav.mx}
\author{Pedro Pasquini~$^2$}\email{pasquini@ifi.unicamp.br}
\author{M. T\'ortola~$^3$}\email{mariam@ific.uv.es}
\author{J. W. F. Valle~$^3$} \email{valle@ific.uv.es, URL:
  http://astroparticles.es/} 

\affiliation{$^1$~\Cinvestav}
\affiliation{$^2$~\campinas}
\affiliation{$^3$~\AddrAHEP}

\begin{abstract}

  We study the capabilities of the short baseline neutrino program at
  Fermilab to probe the unitarity of the lepton mixing matrix. We find the sensitivity to be slightly better than the current one.
  Motivated by the future DUNE experiment, we have also analyzed the potential of an extra liquid Argon near detector in the LBNF
  beamline.  Adding such a near detector to the DUNE setup will substantially improve the current sensitivity on non-unitarity.
  This would help to remove CP degeneracies due to the new complex
  phase present in the neutrino mixing matrix. We also study the
  sensitivity of our proposed setup to light sterile neutrinos for
  various configurations.

  \end{abstract}


\maketitle
\section{Introduction}

The preparation and execution of the DUNE program occupies a central
position in the agenda of neutrino physics experimentation over the
coming decades~\cite{Acciarri:2016ooe,Acciarri:2015uup}.
It is natural that the first phases of the effort will focus on the
short-baseline physics program at Fermilab.
So far the main goal of such an effort has been to confirm or
definitely, rule out the sterile neutrino hints observed in the muon
neutrino beam experiments LSND and MiniBooNE~\cite{Conrad:2013mka}.
While this indeed provides a strong motivation, there are others, of a
more theoretical nature~\cite{Nunokawa:2007qh}, that can further justify the efforts of a
comprehensive short-baseline physics program at Fermilab~\cite{Szelc:2016rjm,Antonello:2015lea}.

Amongst the strong motivations for such a dedicated neutrino program
is the search for short--distance effects associated to neutrino
non-unitarity, as these could possibly shed light on the underlying
seesaw scale associated with neutrino mass
generation~\cite{Schechter:1980gr,Miranda:2016ptb}.
Current limits, as well as future expected sensitivities, have been
discussed
in~\cite{FernandezMartinez:2007ms,Antusch:2014woa,Escrihuela:2016ube,Blennow:2016jkn}~\footnote{
  Astrophysical implications associated with non-unitary evolution of
  solar and supernova neutrinos in matter (and other, more general,
  non-standard interactions) have been widely discussed in the
  literature~\cite{valle:1987gv,nunokawa:1996tg,EstebanPretel:2007yu,Fong:2017gke}.}.
The parameters describing nonunitary neutrino propagation have been
introduced in~\cite{valle:1987gv,nunokawa:1996tg} for the effective
case of two-neutrino mixing. 
A systematic generalized formalism has been presented
in~\cite{Escrihuela:2015wra}, which consistently covers all of the
parameters needed to describe the case of non-unitary three-neutrino
evolution.
One can show that current experiments, involving only electron and
muon neutrinos or anti-neutrinos can be effectively described in
terms of just three new real parameters and one new CP violation
phase.
It has also been shown that this new phase from the seesaw mechanism
brings in a new degeneracy that leads to an important ambiguity in
extracting the "standard" three-neutrino phase
$\delta_{CP}$~\cite{Miranda:2016wdr}. Similar ambiguities in
  the determination of the oscillation parameters can also appear when
  considering the possibility of having light sterile
  neutrinos~\cite{Choubey:2017cba,Agarwalla}. We  discuss the
  potential of our proposed experimental setup for probing this
  scenario as well.  

Recently there has been a lot of interest on the phenomenological
implications of non-unitarity in laboratory searches for neutrino
oscillations~\cite{FernandezMartinez:2007ms,Goswami:2008mi,kopp:2008ds,Antusch:2014woa,Escrihuela:2015wra,Miranda:2016wdr,Escrihuela:2016ube,Blennow:2016jkn,Dutta:2016eks,Dutta:2016czj,Dutta:2016vcc}.
In particular, ways of mitigating the effects of the above discussed
ambiguity for example, by having an additional 20-ton detector in the
TNT2K setup~\cite{Ge:2016dlx} has been addressed in
Ref.~\cite{Ge:2016xya}.
In that case, the main motivation was the use of a cyclotron to
generate a neutrino flux coming from the muon decay at rest ($\mu$DAR)
with a neutrino energy spectrum peaked around 40-50 MeV, to be
detected with a $20$ ton target at $20$~m from the source.
 
Our goal in this paper is to study how the short baseline neutrino
program at Fermilab could help DUNE to break the degeneracy in the
measurement of the CP violation phase associated with the non-unitarity
of the lepton mixing matrix.
Moreover, motivated by the increased interest in Liquid Argon
detectors, we have studied the perspectives of such a detector as a
second near detector for DUNE. We have found that this setup could
substantially improve the sensitivity to non-unitarity parameters.
Indeed, non-unitarity manifests itself mainly as a zero-distance
effect characterizing the effective non-orthonormality of the weak
eigenstate neutrinos~\cite{valle:1987gv,nunokawa:1996tg}.
As a result, improved constraints on the non-unitarity of the neutrino
mixing matrix from short distance measurements would be crucial for
the Long Baseline Neutrino program, as it will help to disentangle the
confusion between the different CP phases appearing in the
neutrino mixing matrix in the non-unitary
case~\cite{Escrihuela:2015wra,Miranda:2016wdr}.

\section{Basic Setup}

In order to introduce the notation, we briefly describe the
effective parameters describing non-unitarity. The structure of
the effective CC weak interaction  mixing matrix is given as
\begin{equation}
  N = N^{NP} U
  \label{eq:Nmat}
\end{equation}
where U is the standard unitary lepton mixing
matrix~\cite{Schechter:1980gr} and the pre-factor matrix $N^{NP}$ is
given as~\cite{valle:1987gv,nunokawa:1996tg}
\begin{equation}\label{eq:NNP}
N^{NP}=\left( \begin{array}{ccc}
\alpha_{11} & 0 & 0 \\
\alpha_{21} & \alpha_{22} & 0 \\
\alpha_{31} & \alpha_{32} & \alpha_{33} \end{array} \right)
\end{equation}
where the diagonal $\alpha_{ii}$ terms are real numbers and the
off-diagonal entries $\alpha_{21},\alpha_{31},\alpha_{32}$ are in
general complex. For a more detailed discussion
see~\cite{Escrihuela:2015wra}.
Constraints on the elements of U arise from global neutrino
oscillation fits~\cite{deSalas:2017kay}.
Laboratory sources of neutrinos are of the electron or muon-types, and
these are described only by the top two rows of the new physics
$N^{NP}$~\cite{Patrignani:2016xqp}. Hence the main parameter probed in
our analysis is $|\alpha_{21}|^2$.

\begin{figure}[t!]
\includegraphics[width=0.6\textwidth,angle=0]{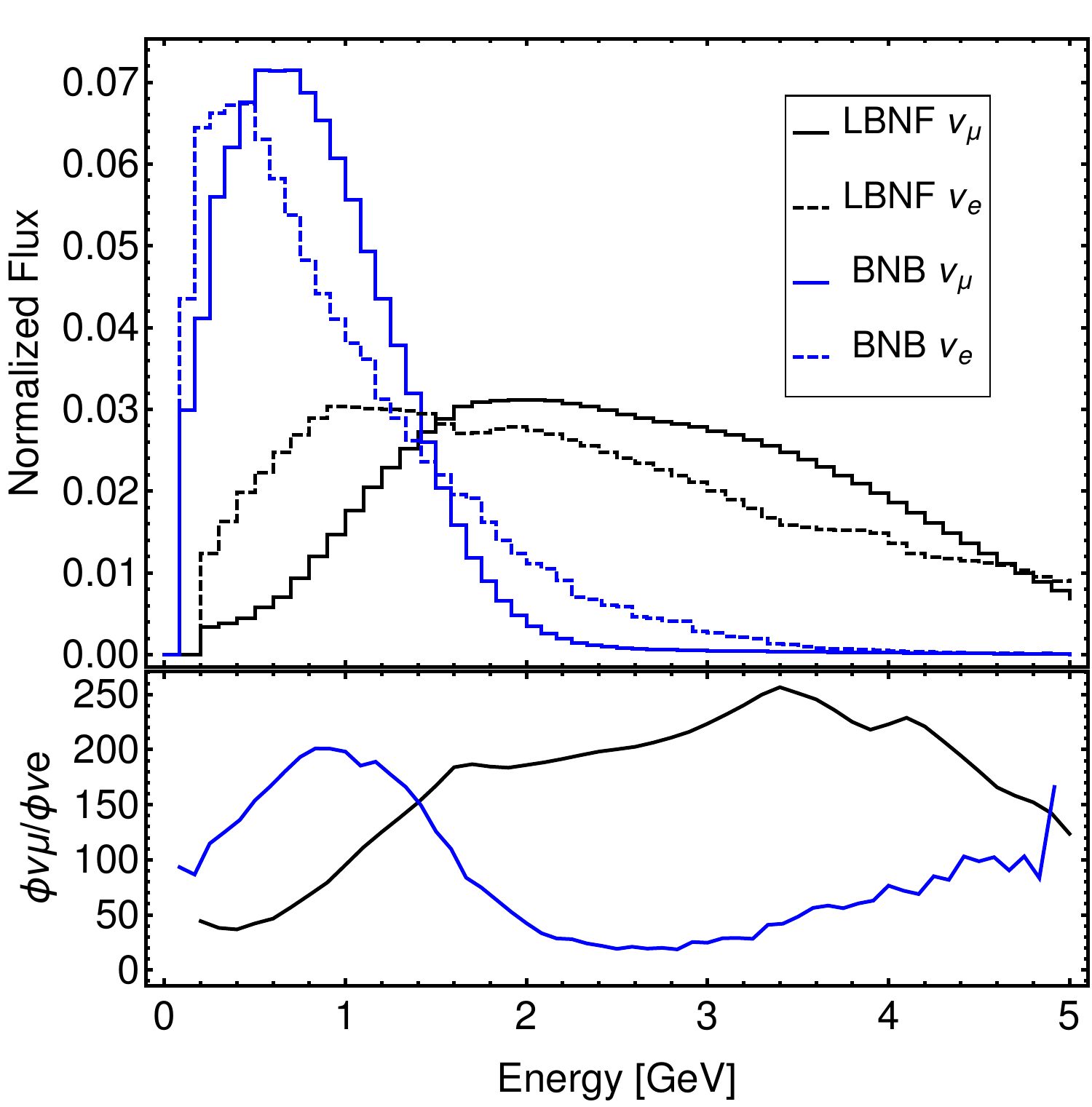}
\caption{Comparison between the normalized neutrino flux from the BNB
  (blue lines) and the LBNF/NUMI beam designed for DUNE (black lines). In
  the upper panel solid (dashed) lines correspond to muon (electron)
  neutrino fluxes.}
\label{fig:fluxes}
\end{figure}

Future short-baseline neutrino experiments aiming to observe light sterile
neutrinos may also be useful to obtain bounds on non-unitary parameters. In
special, non-unitarity predicts a zero distance transition
$\nu_\mu\to \nu_e$~\cite{Escrihuela:2015wra}
\begin{equation}\label{eq:probini}
 P_{\mu e}(L=0)=\alpha_{11}|\alpha_{21}|^2 \, ,
\end{equation}
which can be probed because the initial muon-neutrino fluxes,
$\phi_{\nu_\mu}^0$, in such experiments are much larger than the
  electron neutrino flux contamination, $\phi_{\nu_e}^0$. Thus, at
very short distances from the neutrino source, the number of detected
electron neutrinos $N_e$ is given by
\begin{equation}\label{eq:signal}
 N_e\propto \phi_{\nu_e}^0+|\alpha_{21}|^2\phi_{\nu_\mu}^0 \,.
\end{equation}
In this work we will consider two different sources of neutrinos: the
Booster Neutrino Beam (BNB) and the Long-Baseline Neutrino Facility
(LBNF) beam, also referred to as NUMI beam.  In Fig.~\ref{fig:fluxes}
we provide a comparison of the two Fermilab fluxes, in blue the BNB
flux~\cite{Antonello:2015lea}, featuring an energy peak around 0.6
GeV, and in black the LBNF flux~\cite{Acciarri:2015uup} which peaks
around 2 GeV. The ratio between the muon and electron neutrino fluxe
is typically, $\phi_{\nu_\mu}^0/\phi_{\nu_e}^0\sim 10^2$ in both
cases, which means that one can probe differences in the detected
neutrino spectrum caused by very small values of the non-unitarity
parameter $|\alpha_{21}|^2$.

The current oscillation-only bound on this parameter comes mainly from
the NOMAD experiment~\cite{Astier:2003gs}
\begin{equation}
|\alpha_{21}|^2 < 7.0\times 10^{-4} \, \, \text{at} \, \, 90\% \, \text{C.L.}
\end{equation}
More stringent constraints are obtained when considering charged current
neutrino data. However, one should keep in mind that these limits are
somewhat model--dependent. 

\section{The short-baseline program at Fermilab}

The Fermilab Short Baseline Neutrino Experiment (SBNE) has been
designed to resolve the long-standing puzzle of light sterile
neutrinos~\cite{Abazajian:2012ys}.
The experiment consists of three detectors at different distances: the
Short Baseline Neutrino Detector (SBND), located at 110 m from the
neutrino source, the MicroBooNE detector, at 470 m, and the ICARUS
detector, at 600 m.  Their size and characteristics are described in
Table~\ref{Tab:detec}, summarizing the SBNE
proposal~\cite{Szelc:2016rjm}.
The neutrino source for these three detectors will be the Booster Neutrino Beam (BNB) at Fermilab.
Neutrino beams are generated mainly via by pion, muon and kaon decay.
The pion and kaons are produced by proton collisions and the muons are
generated by the pion decay. Thus, the muon neutrino flux is much
bigger than the electron neutrino flux, as commented above. The BNB
has already operated for several years and its flux is well
understood~\cite{Antonello:2015lea}.  The neutrino beam is obtained
from protons extracted from the Booster accelerator, with around
$5\times10^{12}$~protons per spill hitting a beryllium target 
  with a kinetic energy of $8$~GeV~\cite{Antonello:2015lea}. This
provides a neutrino flux mainly made of muon neutrinos with energies
below $3$~GeV and an energy distribution peaked around 600 MeV. In our
analysis, the main background is the intrinsic electron neutrinos
from the muon and kaon decay.  Of the experiments in
Table~\ref{Tab:detec} MicroBooNE is already running, its detector has
already recorded 3 years of data taking. Thus, our simulation assumes
a total of $6.6\times 10^{20}$ POT for ICARUS and SBND and
$1.32\times 10^{21}$ POT for MicroBooNE.

\begin{table}[t]
 \begin{tabular}{cccccc}
 \hline
  Detector & Total Size & Active Size & Distance & Target & POT\\ \hline \hline
  SBND &  220 t & 112 t & 110 m & Liq. Ar & $6.6\times 10^{20}$\\
  MicroBooNE & 170 t & 89 t & 470 m  & Liq. Ar & $1.32\times 10^{21}$\\
  ICARUS & 760 t & 476 t & 600 m & Liq. Ar &$6.6\times 10^{20}$ \\ \hline
 \end{tabular}
\caption{\label{Tab:detec} Summary of the main features of the SBNE detectors~\cite{Szelc:2016rjm}. }
\end{table}

The Short Baseline Neutrino Experiment (SBNE) at Fermilab contains the
necessary ingredients to observe the non-unitary muon-electron
neutrino transition at short distances. It has an intense flux of muon
neutrino and several detectors located at a short distance that can be
sensitive to zero distance transitions such as
$\nu_\mu\to\nu_e$.

The simulation of the experiment was performed by using the GLoBES
package~\cite{Huber:2004ka,Huber:2007ji}, matching the neutrino fluxes
and detector configurations to those reported in
Ref.~\cite{Szelc:2016rjm}.
In order to include non-unitarity into the GLoBES software we have
modified the build-in numerical calculation of the oscillation
probability using the $S$-Matrix formalism described
in~\cite{Ge:2016xya}. The transition matrix
$S_{\alpha\beta}=\langle \nu_\alpha|e^{-i H L}|\nu_\beta\rangle$ in
the non-unitary case can be calculated by substituting the standard
matter potential by
$V_{\rm NU}= (N N^\dagger) {\rm Diag}[V_{CC}+V_{NC},V_{NC},V_{NC}] (N
N^\dagger)$
 The conventional transition matrix $S^{\rm Unitary}$ of the
  unitary case is given through the relation
\begin{equation}
 S=N^{NP}~S^{\rm Unitary}~\left(N^{NP}\right)^\dagger\,,
\end{equation}
where $N^{NP}$ is the pre-factor describing non-unitarity
  defined in Eq.~(\ref{eq:NNP}).

The expected non-unitarity signal to be searched for would appear as a
change in the total number of events detected with respect to that of
  the unitary case, and also a change in the shape of the electron
neutrino spectrum. Thus, the sensitivity to the parameter
$|\alpha_{21}|^2$ comes from three factors, (i) the relative size
between  $|\alpha_{21}|^2\phi_{\nu_\mu}^0$ and 
  $\phi_{\nu_e}^0$, (ii) the normalization error in the total flux
and (iii) the error in the expected shape of the neutrino flux. Those
are incorporated into the simulation through the $\chi^2$ function
 \begin{equation}
  \chi^2= \sum_{O=1}^3\sum_{i=1}^{N_{\rm bin}} \left(\frac{N_{iO}^{\rm exp}-(1-a-a_{i O})N_{i O}^{\rm th}-(1-b-b_{i O})N_{i O}^{\rm bg}}{\sqrt{N_{iO}^{\rm exp}}} \right)^2+ \chi^2_{\rm SYS} \, ,
  \end{equation}
  with
  \begin{equation}
  \chi^2_{\rm SYS}=\left(\frac{a}{\sigma_a}\right)^2+\left(\frac{b}{\sigma_b}\right)^2+\sum_{O=1}^3\sum_{i=1}^{N_{\rm bin}} \left(\frac{a_{iO}}{\sigma_{sa}}\right)^2+\left(\frac{b_{iO}}{\sigma_{sb}}\right)^2\, ,
 \end{equation}
 where
 $N_{iO}^{\rm exp}\equiv(N_{iO}^{\rm exp})_{\rm sig}+(N_{iO}^{\rm
   exp})_{\rm bkg}$
 is the number of signal and background neutrinos at the $ith$-bin
 expected within the standard unitary 3-neutrino scenario. The
 subscript $O$ runs over the three experiments (SBND, MicroBooNE and ICARUS).  $N_{iO}^{\rm th}$ is the expected number of neutrinos for the transition $\nu_\mu \to \nu_e$ in the non-unitary case and
 $N_{iO}^{\rm bg}$ the expected number of background neutrinos, the
 intrinsic $\nu_e$ from the beam.
Here $\sigma_a$ ($\sigma_b$) is the total neutrino signal (background)
uncertainty and $\sigma_{sa}$ ($\sigma_{sb}$) is the shape signal
(background) uncertainty. All the normalization/shape uncertainties
are taken to be uncorrelated and are incorporated to the simulation
through the minimization of the free parameters $a,b, a_{iO}$ and
$b_{iO}$ for each value of $|\alpha_{21}|^2$.
 
\begin{figure}[t!]
\includegraphics[width=0.45\textwidth,angle=0]{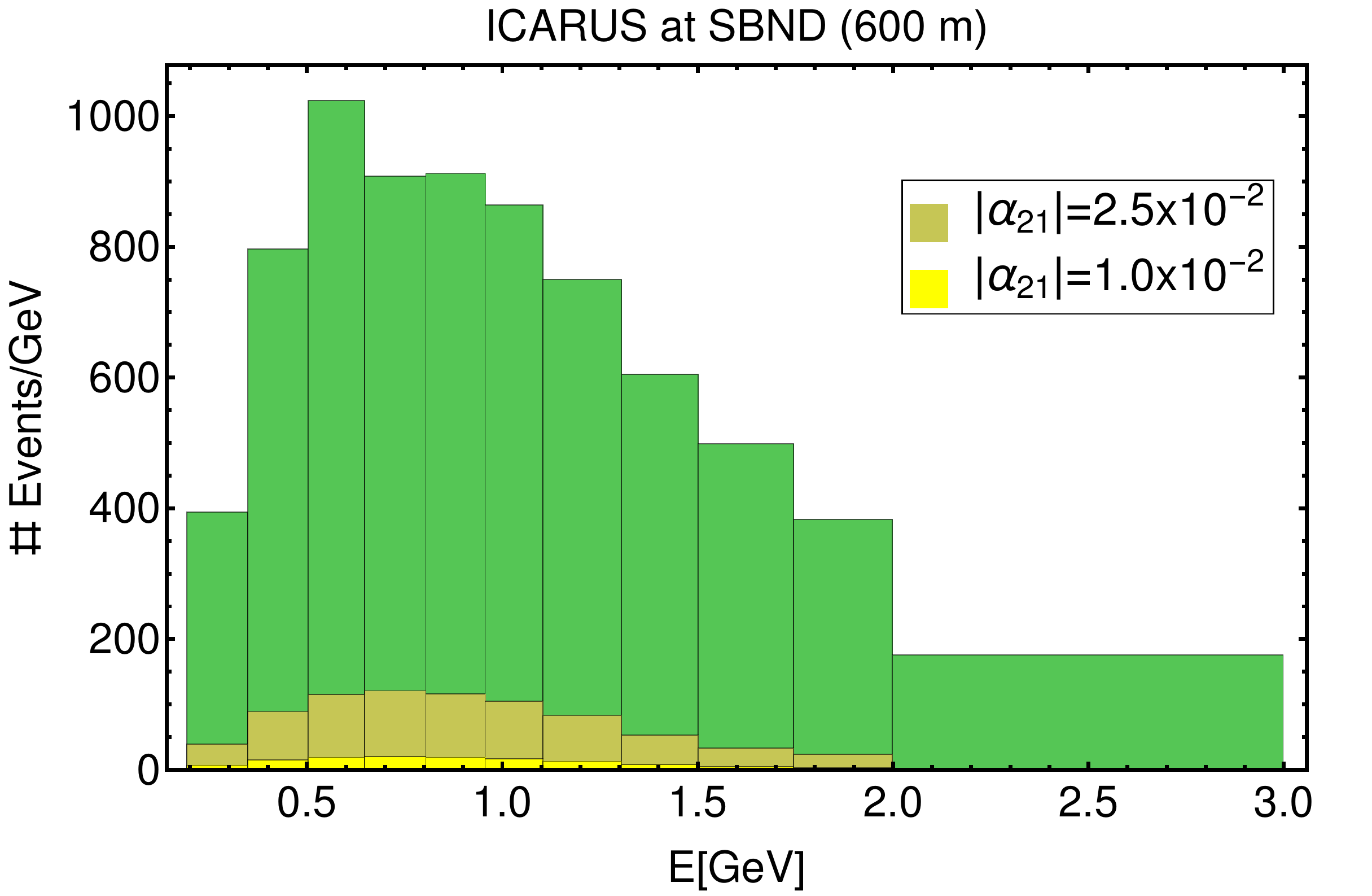}
\includegraphics[width=0.45\textwidth,angle=0]{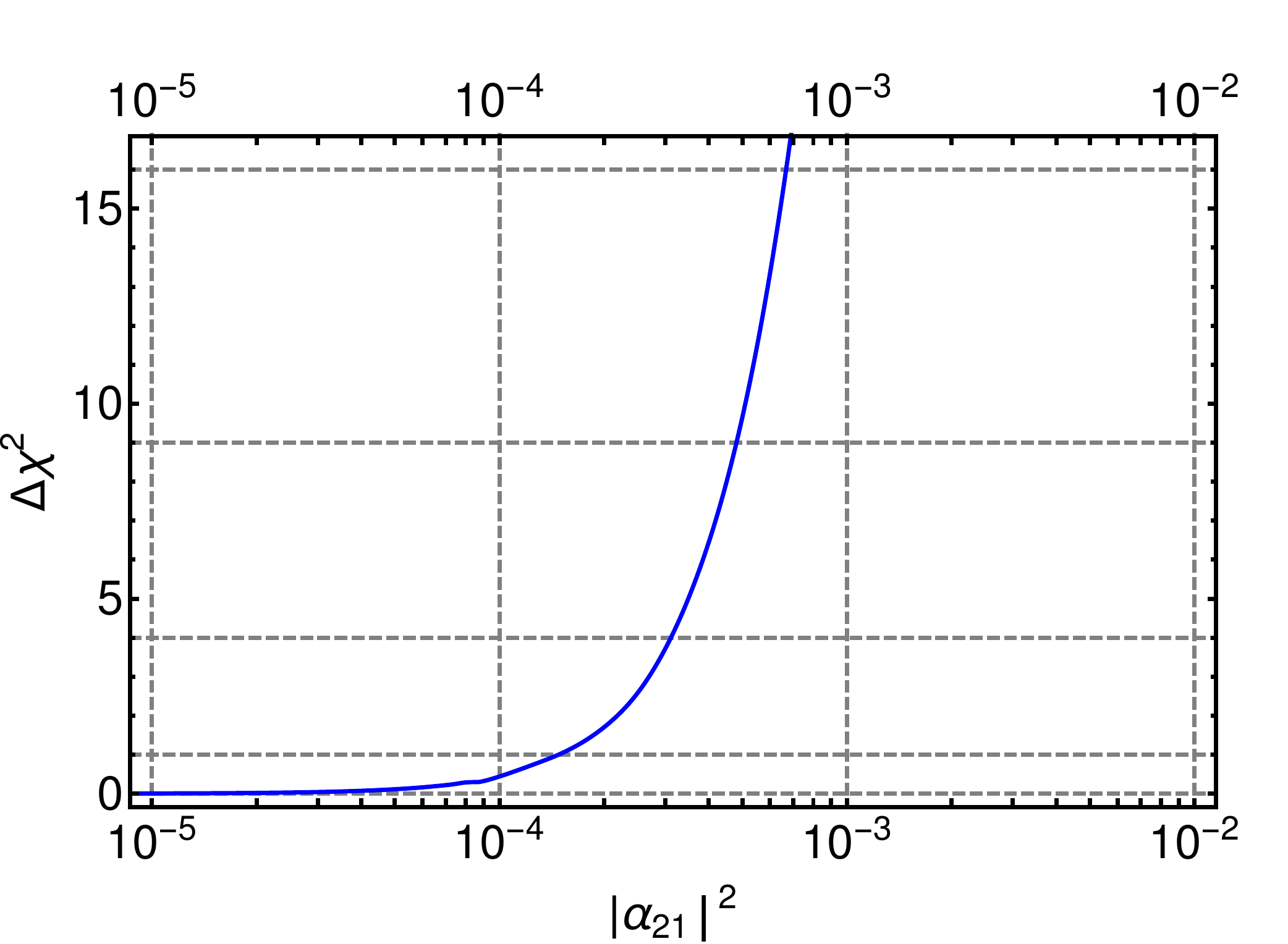}
\caption{{\bf Left:} Number of electron neutrino events (in arbitrary
  units) expected at the ICARUS detector located at the BNB. The green
  histograms show events expected due to the contamination of the
  original neutrino beam, while those expected from the zero-distance
  $\nu_\mu \to \nu_e$ effect due to the non-unitarity signal are in
  dark yellow ($|\alpha_{21}|=2.5\%$) and light yellow (for
  $|\alpha_{21}|=1\%$). {\bf Right:} Expected Sensitivity of SBNE to
  the non-unitarity parameter $|\alpha_{21}|$ assuming a 10\%
  normalization error and a 1\% shape error.  }
\label{fig:SBND_events}
\end{figure}

 The result of the simulation is presented in Fig.~\ref{fig:SBND_events}.
 In the left panel, we present the expected number of electron
 neutrino events at the ICARUS detector from the contamination of the
 original neutrino beam (green) and from the $\nu_\mu \to \nu_e$
 signal associated to non-unitary for $|\alpha_{21}|=2.5\%$ (dark
 yellow) and for $|\alpha_{21}|=1\%$ (light yellow).
 The right panel shows the expected sensitivity of the combined
 analysis of the SBNE experiment (combination of ICARUS, MicroBooNE
 and SBND detectors) to the non-unitarity parameter $|\alpha_{21}|$.
 In our calculations, we have assumed a 10\% normalization error and a
 1\% shape error. With these conditions, the SBNE experiment would
 lead to a sensitivity of $|\alpha_{21}|^2$ at the $3\times10^{-4}$
 level, competitive with current results of non-universality searches.

\section{A second near detector in the LBNF beamline}
We now consider another interesting possibility: the Fermilab
Long-Baseline Neutrino Facility (LBNF) and its near detector program.
As part of the future DUNE experiment~\cite{Acciarri:2015uup},
Fermilab's Main Injector accelerator will be used to produce the LBNF
beamline, providing the highest-intensity neutrino beam in the world.
In this section, we explore the potential of this new neutrino beam as
a probe of the non-unitarity of the lepton mixing matrix.
The DUNE experiment, supplied by the LBNF beam, will already contain a
near detector, located at a distance of approximately $600$~meters.

On the other hand, the ICARUS detector was constructed at CERN and
brought to Fermilab to be assembled as part of the SBNE, as discussed
above.  Here we propose that, after finishing its operation time at
SBNE, the ICARUS detector is transported again so as to be used as a
second near detector in DUNE, sitting at the LBNF neutrino
beamline. As we will now show, this would be very useful in order to
probe non-standard physics.
\begin{figure}[t!]
\includegraphics[width=0.5\textwidth,angle=0]{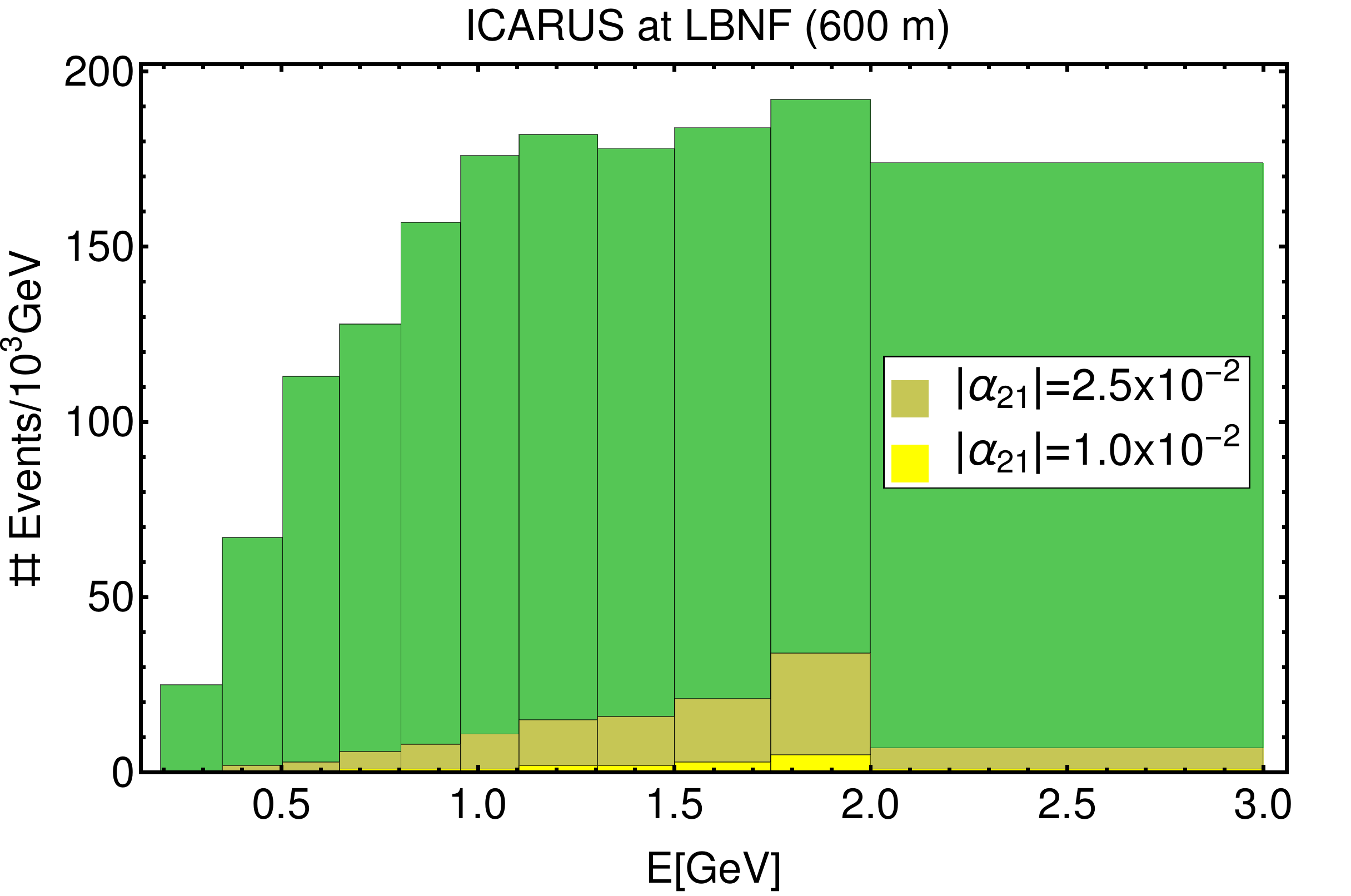}
\caption{Number of electron neutrino events (in arbitrary
  units) expected at the ICARUS detector located at 600 m of the LBNF due to the contamination of
  the original neutrino beam (green). We also show the expected events from 
  a $\nu_\mu \to \nu_e$ conversion due to a non-unitarity signal given by 
  $|\alpha_{21}|=2.5\%$ (dark yellow) and for $|\alpha_{21}|=1\%$
  (light yellow).}
\label{fig:NUMI_events}
\end{figure}

In preparing Figs.~\ref{fig:SBND_events} and \ref{fig:NUMI_events} we
have used the fluxes given in Fig.~\ref{fig:fluxes}.  The latter gives
the fluxes needed to estimate the expected event number at the ICARUS
detector placed at the LBNF beamline used as the neutrino source.
Notice that the number of events is much bigger at LBNF than in the
BNB setup considered in Fig.~\ref{fig:SBND_events}.
Nevertheless, the ICARUS detector is not optimized for the LBNF beam,
since it was designed for a less energetic beam, like the BNB beam,
with neutrino energies ranging from $0$ to $3$ GeV with a peak around
0.6 GeV. The LBNF beamline, on the other hand, contains neutrinos from
0 to 5 GeV and peaks at 2 GeV.

In order to take into account the above features, we have considered
three possible configurations for the proposed second near detector at
the LBNF beam:
\begin{enumerate}
 
\item ICARUS at LBNF: This is exactly the ICARUS detector of the SBNE,
  working with energies between 0 to 3 GeV, located at the LBNF
  beamline. 
 
\item ICARUS+ at LBNF: Again the same ICARUS detector of the SBNE,
  working with an extended energy window between $0$ and $5$ GeV and located at the LBNF beamline. In this case we have added an extra energy bin to the experiment simulation, corresponding to energies from 3 to 5 GeV. For this extra bin, we have assumed the same efficiency as in the previous energy bin.
 
\item A protoDUNE-like detector~\cite{Manenti:2017txy}: We have
    assumed the standard single phase DUNE Liquid Argon far detector
    configuration, with the proposed efficiency, bin size, etc and
    with an active mass corresponding to a $450$~ton detector, as
    considered for the ProtoDUNE-SP detector.

\end{enumerate}   
There is also a $300$~ton detector possibility, the Dual-Phase
protoDUNE detector. Although has a smaller mass, this would employ a
combination of liquid and gas Argon that may present an advantage over
the standard protoDUNE Single Phase detector described
above. Nevertheless, in our analysis, we will consider the simpler case
of the single-phase detector, as the expected performance and design
for the dual phase detector are not yet settled.
A summary of these detectors can be found in
Table~\ref{Tab:detec2}. Although the final design has not been fixed
yet, the protoDUNE configuration described above is much more similar
to what the DUNE near detector will be.

\begin{table}[t]
 \begin{tabular}{ccccc}
 \hline
  Detector &  Active Size & Distance & E range (GeV) & Target \\ \hline \hline
  ICARUS &  476 t & 600 m & 0 to 3 & Liq. Argon\\
  ICARUS+ &  476 t & 600 m & 0 to 5 & Liq. Argon\\
  protoDUNE-SP &  450 t & 600 m & 0 to 5 & Liq. Argon\\
\hline
 \end{tabular}
\caption{\label{Tab:detec2} {Proposals for a second near detector in DUNE. }}
\end{table}

\subsection{Sensitivity to Non-Unitarity at an LBNF near detector}
\begin{figure}[t!]
\includegraphics[width=0.65\textwidth,angle=0]{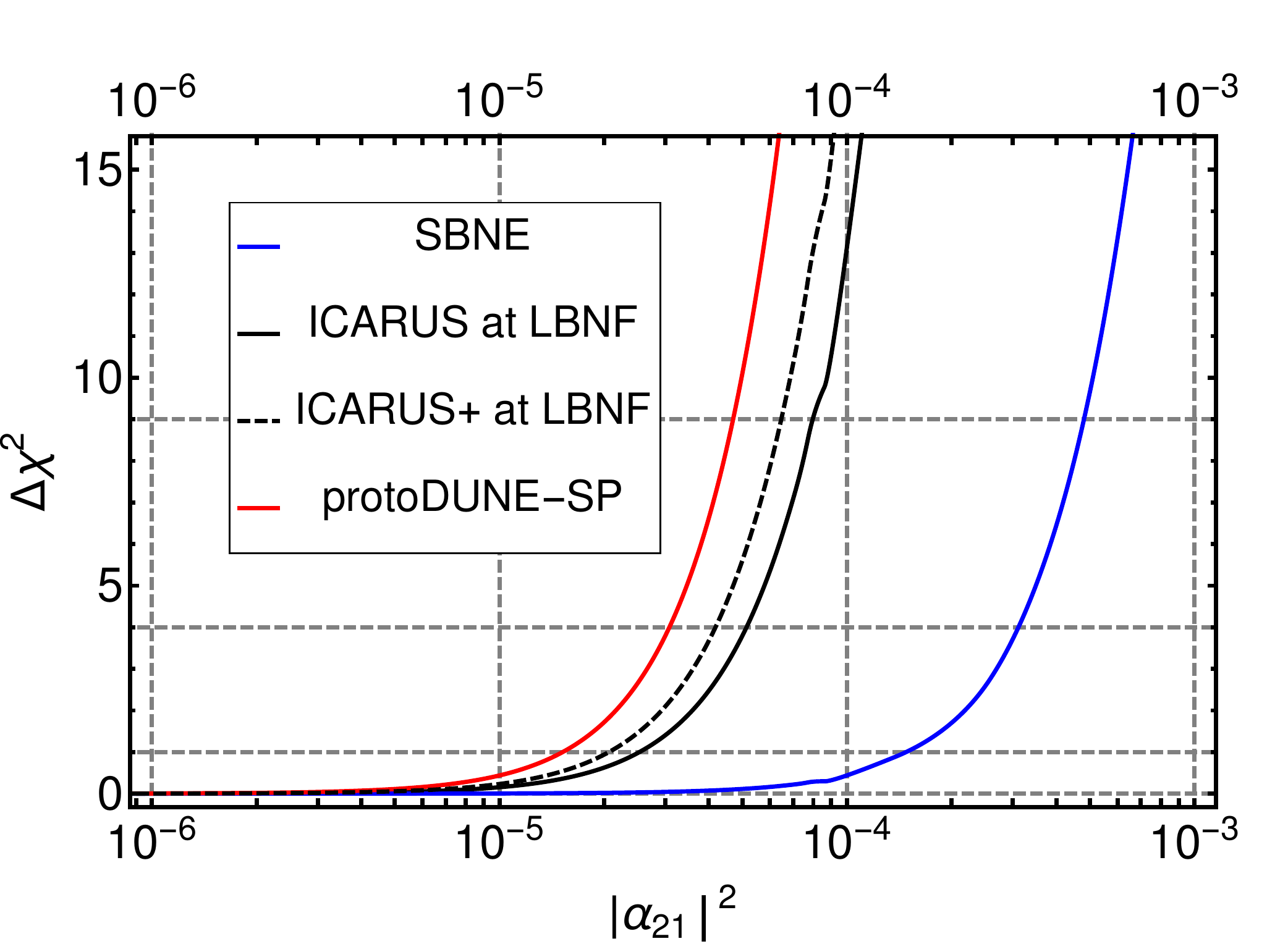}
\caption{Sensitivity of each configuration assumed: SBN experiment
  (blue), ICARUS at LBNF (black-solid), ICARUS+ at LBNF (black-dashed)
  and protoDUNE-SP (red). All of them are assumed to be located
  at 600 m from the neutrino source and running for 3.5 years in the
  neutrino and 3.5 in the anti-neutrino mode.}
\label{fig:ICARUS_NUMI_sensi}
\end{figure}

The resulting sensitivity of each of the detectors proposed in Table
\ref{Tab:detec2} is plotted in Fig.~\ref{fig:ICARUS_NUMI_sensi}. Here
we are assuming 10\% normalization error and 1\% shape error.  One
sees that thanks to the high statistics of LBNF beam, the expected
sensitivities in these cases are quite promising.  Indeed, these
configurations result in a substantial improvement of one order of
magnitude of the sensitivity to $|\alpha_{21}|^2$.

\begin{figure}[t!]
\includegraphics[width=0.9\textwidth,angle=0]{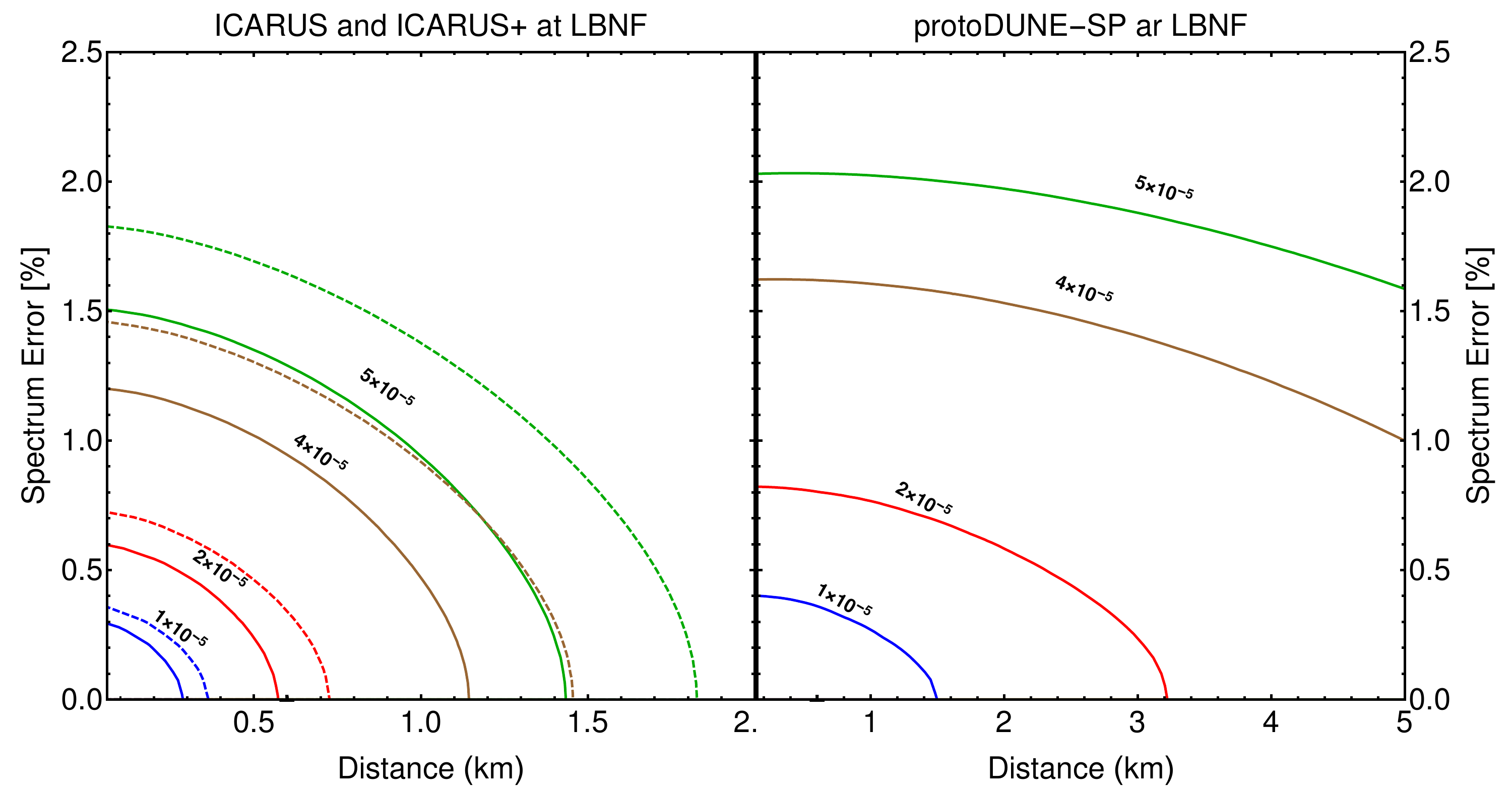}
\caption{  {\bf Left:} 90\% C.L. sensitivity to $|\alpha_{21}|$
  for ICARUS (solid line) and ICARUS+ (dashed line) for various
  combinations of the baseline and the spectrum error. {\bf Right:}
  90\% C.L. protoDUNE-SP sensitivity for various combinations of
  baseline and spectrum error. Lines correspond to
  $|\alpha_{21}|^2<10^{-5}$ (blue), $|\alpha_{21}|^2<2\times 10^{-5}$
  (red), $|\alpha_{21}|^2<4\times10^{-5}$ (brown) and
  $|\alpha_{21}|^2<5\times10^{-5}$ (green).}
\label{fig:ICARUS_NUMI}
\end{figure}

The biggest drawback of these experiments is the requirement of
knowing precisely the shape of the neutrino flux with high
precision. The DUNE collaboration will predict the neutrino flux by
measuring the muons and hadron-production responsible for the neutrino
beam~\cite{Acciarri:2016ooe}. In Fig.~\ref{fig:ICARUS_NUMI} we present
the sensitivity on $|\alpha_{21}|^2$ at 90\% C. L. for various
combinations of the baseline and the assumed uncertainty in the
neutrino spectrum. Notice that the spectrum error limits the maximum
attainable sensitivity on $|\alpha_{21}|^2$. For example, the
protoDUNE configuration cannot reach
$|\alpha_{21}|^2<2.5\times10^{-5}$ if the spectrum is not known up to
a $1\%$ precision.

\begin{figure}[H]
\centering
\includegraphics[width=0.6\textwidth,angle=0]{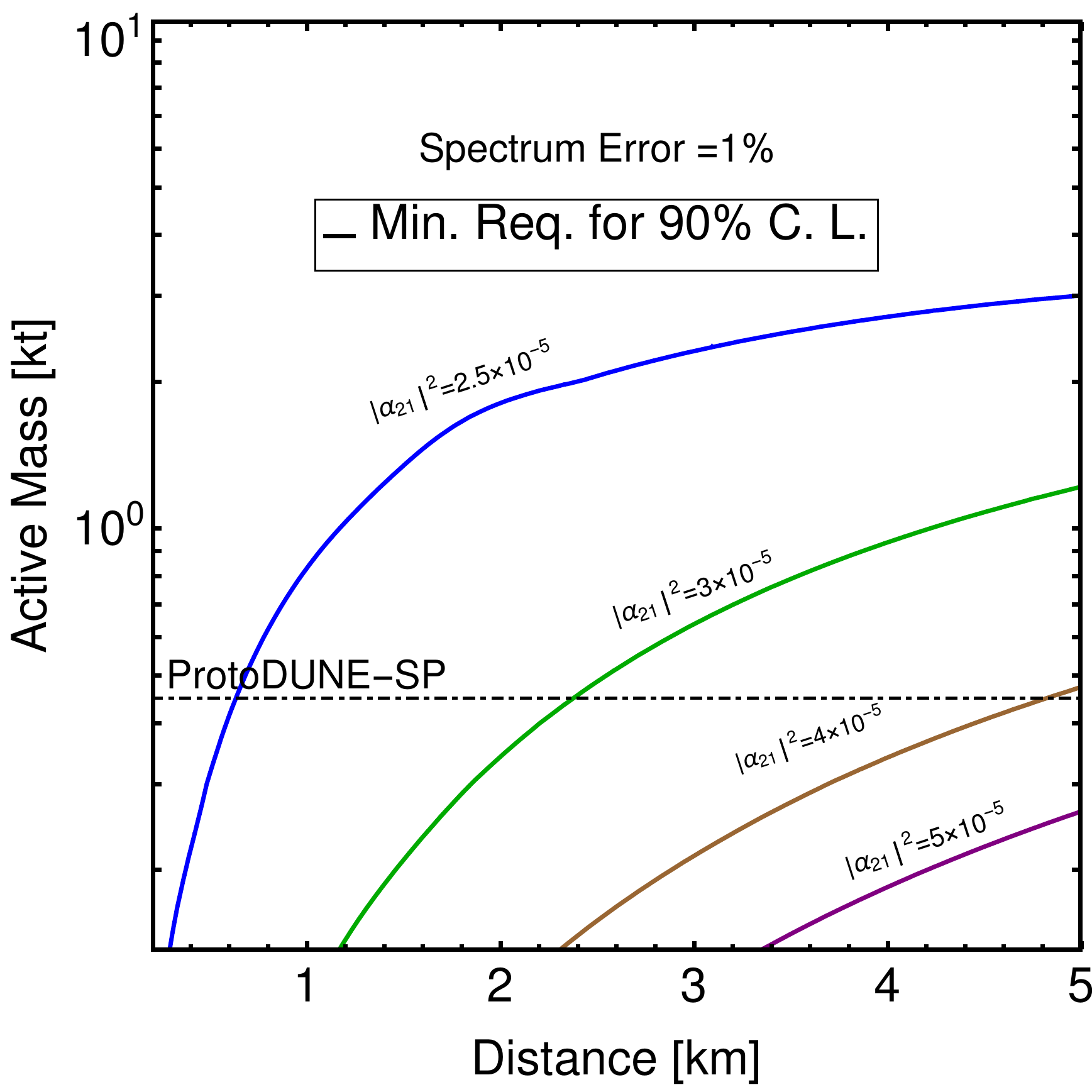}
\caption{ Minimum detector distance and active mass requirements
  to attain the indicated $90\%$ C.L. sensitivity for different
  values of $|\alpha_{21}|^2$, assuming an spectrum error of $1\%$.}
\label{fig:minimum}
\end{figure}

The discussion of Fig.~\ref{fig:ICARUS_NUMI} can also be extended by
considering the impact of the different detector sizes and
distances. The results of this analysis are displayed in
Fig.~\ref{fig:minimum}.
There we have plotted the minimum requirements for obtaining a $90\%$
C. L. bound for different values of $|\alpha_{21}|^2$ assuming a
spectrum error of $1\%$. This figure clearly illustrates that, as
expected, the smaller the detector, the closer it should be put in
order to obtain a good sensitivity. Nevertheless, notice that even
with a very large detector size, one can not improve the ``ultimate''
precision on $|\alpha_{21}|^2<2.5\times10^{-5}$ for the assumed $1\%$
spectrum precision.
\subsection{Sensitivity to light Sterile Neutrinos at an LBNF near detector}

Here we focus on the short-baseline capabilities of Fermilab concerning
the sensitivity to light sterile neutrinos in the eV range.
The LNBF near detector will be located at around 600 m from the beam
source. This opens up a possibility to probe not only zero-distance
effects, but also effects that change with energy and distance, such
as those associated with a light sterile neutrino.  
 In fact, one could use one (or several) near
  detector(s) at the LBNF beamline in conjunction with
beamline spectrum measurement to probe light sterile neutrinos. 
The possibility of probing sterile neutrino oscillations using a near
detector in the DUNE experiment has already been considered in
Ref.~\cite{Choubey:2016fpi}. Although the appearance channel is
  in general the most sensitive, here we notice that the sensitivity
  to light sterile neutrinos in the disappearance channel may be
  substantially improved provided the uncertainty in the shape of the  
  neutrino spectrum  is good enough. To illustrate this point we consider different values of the spectrum error, 
  as well as the possibility of combining two different near
  detectors in the LBNF beamline.
We also pay especial attention to the effect of the distance from the source to the detector.
%
For definiteness, we assume a 3+1 neutrino scheme, since the symmetric
2+2 schemes~\cite{peltoniemi:1993ss,peltoniemi:1993ec} are ruled out
by the solar and atmospheric neutrino oscillation
data~\cite{Maltoni:2004ei,Kopp:2013vaa,Gariazzo:2017fdh}.
In the usual framework, the standard oscillation paradigm contains three  active
neutrinos that oscillate to one another. For a neutrino beam of energy
around  $2.5$ GeV, a baseline of around $10^{3}$ km would be
required for the oscillation to take place.
Nevertheless, the existence of one (or several) sterile neutrinos with
mass-squared differences $\Delta m_{n1}^2$ , with $n>3$, around the
eV$^2$ scale would potentially give rise to oscillations in the scale
of hundreds of meters. 
\begin{figure}[t]
\centering
\includegraphics[width=0.99\textwidth,angle=0]{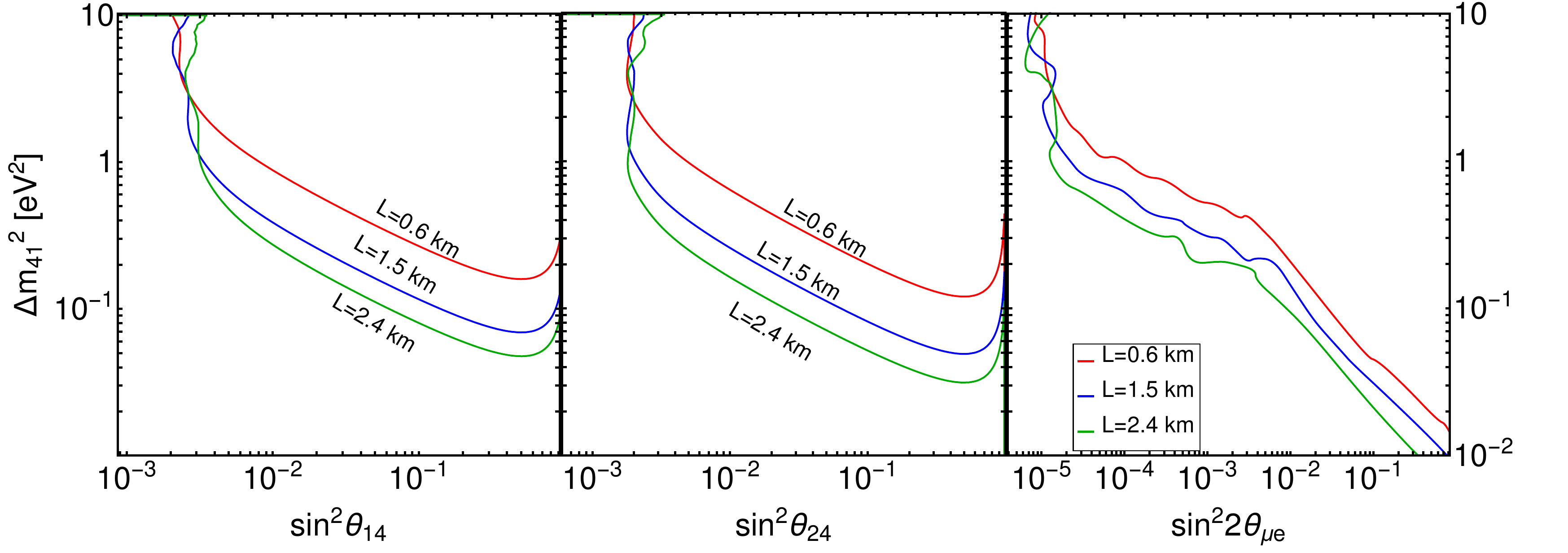}
\caption{ProtoDUNE-SP sensitivity at 90\% CL to the 3+1 neutrino
  scheme for three baselines at the LBNF beam: $L=0.6$ km (red),
  $L=1.5$ km (blue) and $L=2.4$ km (green). {\bf Left:}
  $\sin^2\theta_{14}$ versus $\Delta m_{41}^2$   {\bf Center:}
 $\sin^2\theta_{24}$ versus $\Delta m_{41}^2$  and {\bf Right:}
  $\sin^2 2\theta_{\mu e}$ versus $\Delta m_{41}^2$.  A $1\%$ spectrum
  error is assumed in all cases.}
\label{fig:proposals}
\end{figure}

\begin{figure}[t]
\centering
\includegraphics[width=0.99\textwidth,angle=0]{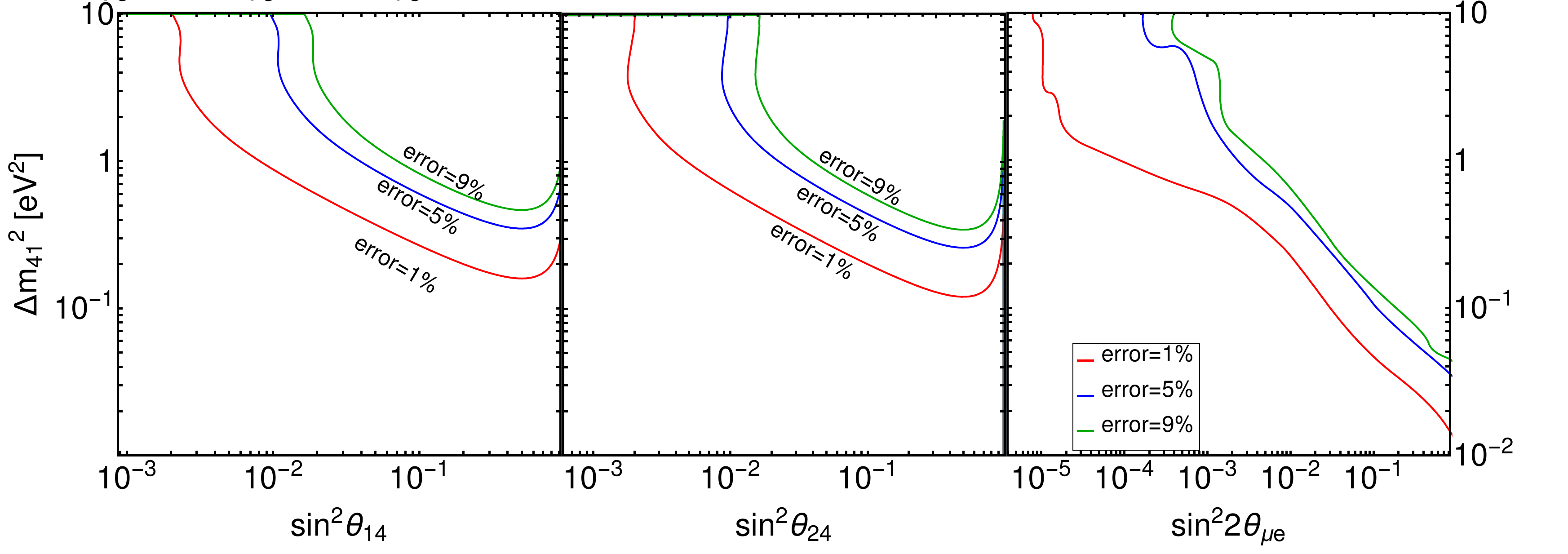}
\caption{ Effect of the spectrum error on the at 90\% CL
  ProtoDUNE-SP sensitivity to the 3+1 neutrino scheme for three
  different cases: $1\%$ (red), $4\%$ (blue) and $9\%$ (green).
  {\bf Left:} $\sin^2\theta_{14}$ versus $\Delta m_{41}^2$   {\bf Center:}
 $\sin^2\theta_{24}$ versus $\Delta m_{41}^2$  and {\bf Right:}
  $\sin^2 2\theta_{\mu e}$ versus $\Delta m_{41}^2$. We have
  assumed a baseline of $0.6$ km.}
\label{fig:proposals2}
\end{figure}

\begin{figure}[t]
\centering
\includegraphics[width=0.99\textwidth,angle=0]{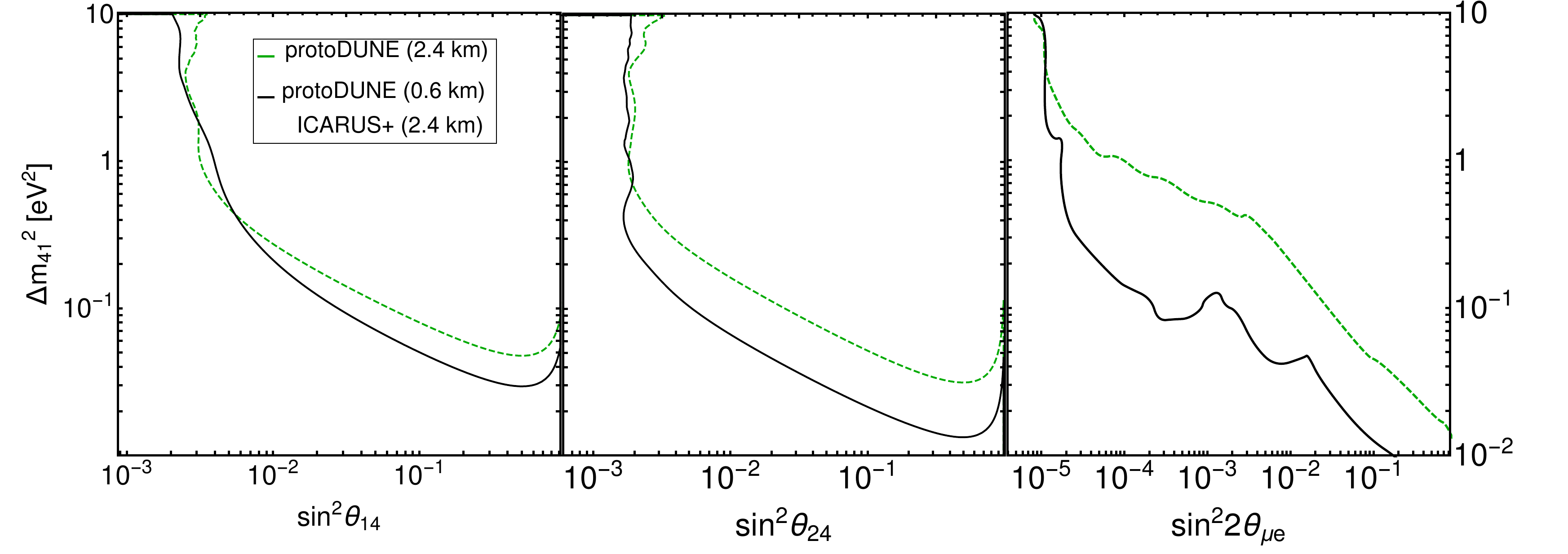}
\caption{ The LBNF near detectors at 90\% C.L. sensitivity to the
  3+1 neutrino scheme is given in black for the combination of
  protoDUNE-SP at 0.6 km and ICARUS+ at 2.4 km. The Dashed-Green curve
  shows the result for the protoDUNE-only case at 2.4 km from the
  LBNF. {\bf Left:}
  $\sin^2\theta_{14}$ versus $\Delta m_{41}^2$   {\bf Center:}
 $\sin^2\theta_{24}$ versus $\Delta m_{41}^2$  and {\bf Right:}
  $\sin^2 2\theta_{\mu e}$ versus $\Delta m_{41}^2$.  A
  $1\%$ spectrum error is assumed in all cases.}
\label{fig:proposals3}
\end{figure}

In Fig.~\ref{fig:proposals} we illustrate the effect of the baseline
on the sensitivity to the 3+1 neutrino scheme of the protoDUNE-SP
detector located at the LBNF beamline as proposed previously.
We plot the expected sensitivity in the
  $\sin^2\theta_{14}$--$\Delta m^2_{41}$,
  $\sin^2\theta_{24}$--$\Delta m^2_{41}$  and
  $\sin^22\theta_{\mu e}$--$\Delta m^2_{41}$
  planes, where $\sin^2 2\theta_{\mu e}=4|U_{e4}|^2|U_{\mu4}|^2$. We consider different
baselines and assume a $1\%$ spectrum error.  The experiment is
  not sensitive to $\theta_{34}$.  Fig.~\ref{fig:proposals2} shows
the impact of the spectrum error measurement for a 0.6 km baseline
protoDUNE-SP detector.
 In contrast to usual sterile neutrino searches, the DUNE
  experiment has a clear advantage, since it is sensitive to three
  channels: $\nu_e \to \nu_e$, $\nu_\mu \to \nu_\mu$ and
  $\nu_\mu \to \nu_e$.  This allows one to constrain the values of
  $\theta_{14}$ and $\theta_{24}$ separately. In order to see this
  quantitatively, we have estimated the sensitivity of each
    disappearance channel in
  Figs.~\ref{fig:proposals},~\ref{fig:proposals2}
  and~\ref{fig:proposals3}. In the left panel of these figures
    we have focused on the electron neutrino disappearance channel,
    setting $\theta_{24}=0$, while the central panel assumes
  $\theta_{14}=0$ and  shows the sensitivity to muon neutrino
    disappearance alone. The combined sensitivity on the
    sterile neutrino parameters coming from the disappearance channels
    and the appearance channel $\nu_\mu \to \nu_e$ is shown at the
  right panel, and has also been discussed in~\cite{Choubey:2016fpi}.

The usual configuration of a sterile neutrino experiment consists on a
very near detector that supplies the spectrum measurement of the
beamline. This could be accomplished by using the protoDUNE-like near
detector at 0.6 km and the ICARUS detector at 2.4 km. This
configuration improves the sensitivity to probe the 3+1 parameter
space as can be seen on Fig.~\ref{fig:proposals3}. 
The green line corresponds to the sensitivity curve of the
protoDUNE-only configuration located at 2.4 km from the neutrino
source, while the black line corresponds to the combination of ICARUS+
at 2.4 km and protoDUNE at 0.6 km. Notice that in general the
combination of detectors improves the sensitivity since protoDUNE
would act as a near detector for ICARUS+, providing a good estimate
for the shape of the neutrino flux. Nevertheless, at
$\Delta m^2\sim 1$ eV$^2$ the sensitivity of protoDUNE alone is
slightly better, as 2.4 km is the optimal baseline for neutrino
oscillations with mass squared splitting around 1eV$^2$ and protoDUNE
is a detector optimized for the LBNF flux.

\vspace*{0.5cm}

Comparing these results with other sensitivity studies performed
  in the literature, for experiments such as
  Hyper-Kamiokande~\cite{Kelly:2017kch,Tang:2017khg} or  MINOS+~\cite{Tzanankos:2011zz,Adamson:2017uda}, one can see that the
  experimental setups proposed here look very promising indeed,
  especially for constraining  $\sin^2 2\theta_{\mu e}$.

\section{Discussion and conclusion}
We have studied the capabilities of the short baseline neutrino
program at Fermilab as a probe of the unitarity of the lepton mixing
matrix. In particular, we have analyzed in this case the 
sensitivity to the so-called zero distance effect. 
We have found that the sensitivity is slightly better than the current
one from oscillation experiments such as NOMAD, especially when the
analyses of the three upcoming detectors are combined, as shown in
Fig.~\ref{fig:SBND_events}.
Motivated by the future DUNE experiment, we have also analyzed the
potential of different liquid Argon near detectors located in the LBNF
beamline.  We have found that the addition of such a near detector
 to the DUNE setup can substantially
improve the current sensitivity on non-unitarity parameters.
Fig.~\ref{fig:ICARUS_NUMI_sensi} illustrates the improvement in the
sensitivity to unitarity violation that can be achieved in this case.
Such improvement would help to remove the degeneracies associated with
the search for CP violation at DUNE, coming from the new complex phase
present in the non-unitary neutrino mixing
matrix~\cite{Miranda:2016wdr}.
For completeness, we have also analyzed in detail how the sensitivity
changes for different configurations of baseline, mass, and systematic
errors, as summarized in Figs.~\ref{fig:ICARUS_NUMI} and
\ref{fig:minimum}.\\[-.2cm]

We have also commented on the use of such a DUNE near detector, such
as a probe for light sterile neutrinos. 
We have studied the sensitivity of various configurations of
baselines and errors (see Figs.~\ref{fig:proposals} and
\ref{fig:proposals2}). We have also studied the case (Fig. 9) of an
array of two near detectors located at $0.6$ and $2.4$~km that could
probe the $\Delta m^2\sim$ 1 eV$^2$ region both for $\theta_{14}$ and
$\theta_{24}$.
The impact of having a second near detector is especially
  visible in the expected sensitivity to $\sin^2 2\theta_{\mu e}$, plotted in the right panel of Fig.~\ref{fig:proposals3}.
\\[-.2cm]

Finally, an LBNF near detector can also probe neutrino non-standard
interactions (NSI).
Such NSI are generically expected in neutrino mass generation schemes,
not necessarily of the seesaw type~\cite{Boucenna:2014zba}. 
Indeed, the sensitivity to NSI in the DUNE far detector has already been
discussed in
Refs.~\cite{Masud:2017bcf,deGouvea:2015ndi,Coloma:2015kiu,Liao:2016orc}.
Here we stress that such interactions also lead to an effective
non-unitarity-like zero--distance effect, ideal to be probed at a near
detector.
For the case of short--baseline neutrino experiments, matter effects
in the neutrino propagation are irrelevant, and therefore the
experiments are only sensitive to NSI at the neutrino production or
detection processes.
One can parametrize the charged current NSI at the neutrino source (s)
and detection (d) in terms of two $3\times3$ matrices: $\epsilon^s$
and $\epsilon^d$~\cite{Farzan:2017xzy} that modify the oscillation
probability to~\cite{Blennow:2016jkn}
\begin{equation}
 P_{\alpha\beta}=|[(1+\epsilon^d)S(1+\epsilon^s)]_{\beta \alpha}|^2,
\end{equation}
where $S$ is the propagation matrix. The limit $\epsilon^{a}\to 0$,
with $a=s,d$, restores the standard oscillation result. The analogue
zero--distance effect corresponding to Eq.~(\ref{eq:signal}) becomes
\begin{equation}\label{eq:signal_NSI}
 N_e\propto |(1+\epsilon^s_{ee})(1+\epsilon^d_{ee})+\epsilon^d_{e\mu }\epsilon^{s}_{\mu e}|^2\phi_{\nu_e}+|(1+\epsilon^s_{ee})\epsilon^d_{e\mu }+(1+\epsilon^d_{\mu\mu})\epsilon^s_{e\mu}|^2\phi_{\nu_\mu}
\end{equation}
Therefore, all the analyses obtained before can be extended to cover
this case as well, by substituting $|\alpha_{21}|^2$ by the quantity,
\begin{equation}\label{eq:signal_NSI2}
|\alpha_{21}|^2\rightarrow \frac{|(1+\epsilon^s_{ee})\epsilon^d_{e\mu}+(1+\epsilon^d_{\mu\mu})\epsilon^s_{e\mu}|^2}{|(1+\epsilon^s_{ee})(1+\epsilon^d_{ee})+\epsilon^d_{e\mu }\epsilon^{s}_{\mu e}|^2}\approx |\epsilon^d_{e\mu }+\epsilon^s_{e\mu}|^2
\end{equation}
Notice that the experiment becomes blind to NSI in the special case
$\epsilon^d_{\mu e}\approx -\epsilon^s_{\mu e}$.\\[-.2cm]

In summary, our main point in this paper has been to stress the
  importance of probing short distance physics through the use of near detectors in DUNE. We have illustrated the physics that can be probed in several different configurations.
  In order to bring the issue to the experimental agenda we have
  proposed idealized benchmarks and determined their physics reach.
  Our results should trigger discussion in the community and help
  choose an optimized and realistic option. Dedicated scrutiny will be
  needed in order to design the ultimate setup to be chosen, in view
  of its physics interest as well as technical feasibility.  \vfill

\begin{acknowledgments}

  The authors would like to thank G. V. Stenico for providing the SBND
  GloBES code for the detectors.
  Work supported by the Spanish grants FPA2017-85216-P and
  SEV-2014-0398 (MINECO), PROMETEOII/2014/084 and GV2016-142 grants
  from Generalitat Valenciana.
  MT is also supported a Ram\'{o}n y Cajal contract (MINECO). OGM
  was supported by CONACyT and SNI (Mexico).
  P. P. thanks the support of FAPESP-CAPES funding grant 2014/05133-1,
  2014/19164-6 and 2015/16809-9 Also the partial support from FAEPEX
  funding grant, No 2391/17.

\end{acknowledgments}

\bibliographystyle{apsrev} 
\providecommand{\url}[1]{\texttt{#1}}
\providecommand{\urlprefix}{URL }
\providecommand{\eprint}[2][]{\url{#2}}

\end{document}